\documentstyle[11pt,iaus212,twoside,epsf]{article}

\def\lsim{\mathrel{\rlap{\lower 4pt \hbox{\hskip 1pt $\sim$}}\raise 1pt \hbox
        {$<$}}}
\def\gsim{\mathrel{\rlap{\lower 4pt \hbox{\hskip 1pt $\sim$}}\raise 1pt \hbox
        {$>$}}}
\newcommand{\ms}{$M_\odot$}
\newcommand{\Nifs}{$^{56}$Ni}

\newcommand{\Msun}{$M_\odot$}

\def\edcomment#1{\iffalse\marginpar{\raggedright\sl#1\/}\else\relax\fi}
\reversemarginpar

\begin{document}
\vspace*{1cm}
\title{Hypernovae and their Nucleosynthesis
}
\author{
Ken'ichi Nomoto$^{1}$, Keiichi Maeda$^{1}$, Hideyuki Umeda$^{1}$ \\ 
Takuya Ohkubo$^{1}$, Jingsong Deng$^{1}$, Paolo Mazzali$^{2}$}
\affil{
$^{1}$Department of Astronomy, School of Science, University of Tokyo,
 7-3-1 Hongo, Bunkyo-ku, Tokyo 113-0033, Japan\\
$^{2}$Osservatorio Astronomico, Via Tiepolo, 11, 34131 Trieste, Italy
}

\begin{abstract}

We review the characteristics of nucleosynthesis in 'Hypernovae',
i.e., core-collapse supernovae with very large explosion energies ($
\gsim 10^{52} $ ergs).  The hypernova yields show the following
characteristics: 1) The mass ratio between the complete and incomplete
Si burning regions is larger in hypernovae than normal supernovae.  As
a result, higher energy explosions tend to produce larger [(Zn, Co,
V)/Fe] and smaller [(Mn, Cr)/Fe], which could explain the trend
observed in very metal-poor stars.  2) Because of enhanced
$\alpha$-rich freezeout, $^{44}$Ca, $^{48}$Ti, and $^{64}$Zn are
produced more abundantly than in normal supernovae.  The large [(Ti,
Zn)/Fe] ratios observed in very metal poor stars strongly suggest a
significant contribution of hypernovae.  3) Oxygen burning takes place
in more extended regions in hypernovae to synthesize a larger amount
of Si, S, Ar, and Ca ("Si"), which makes the "Si"/O ratio larger.  The
abundance pattern of the starburst galaxy M82 may be attributed to
hypernova explosions.  We thus suggest that hypernovae make important
contribution to the early Galactic (and cosmic) chemical evolution.

\end{abstract}

\section{Introduction}

 One of the most interesting recent developments in the study of
supernovae (SNe) is the discovery of some very energetic Supernovae
(SNe), whose kinetic energy (KE) exceeds $10^{52}$\,erg, about 10
times the KE of normal core-collapse SNe (hereafter $E_{51} =
E/10^{51}$\,erg).  Type Ic supernova (SN~Ic) 1998bw was probably
linked to GRB 980425 (Galama et al. 1998), thus establishing for the
first time a connection between gamma-ray bursts (GRBs) and the
well-studied phenomenon of core-collapse SNe.  However, SN~1998bw was
exceptional for a SN~Ic: it was as luminous at peak as a SN~Ia,
indicating that it synthesized $\sim 0.5$ \Msun\ of \Nifs, and its
KE was estimated at $E \sim 3 \times 10^{52}$ erg (Iwamoto et
al. 1998; Woosley et al. 1999).  Because of its large KE, SN~1998bw
was called a ``Hypernova (HN)".

Subsequently, other ``hypernovae" of Type Ic have been discovered or
recognized, such as SN~1997ef (Iwamoto et al. 2000; Mazzali, Iwamoto,
\& Nomoto 2000), SN~1997dq (Matheson et al. 2001), SN~1999as (Knop et
al. 1999; Hatano et al. 2001), and SN~2002ap (Mazzali et al. 2002).
Another possible hypernovae, although of Type IIn, were SNe~1997cy
(Germany et al. 2000; Turatto et al. 2000) and 1999E (Rigon et
al. 2002).  Figure 1 shows the near-maximum spectra and the absolute
V-light curves of Type Ic hypernovae.  These hypernovae span a wide
range of properties, although they all appear to be highly energetic
compared to normal core-collapse SNe.  SN 1999as is the most luminous
supernova ever discovered, reaching a peak magnitude $M_{\rm V} < -
21.5$, while the brightness of SN 2002ap appears to be similar to that
of normal core collapse SNe.

In the following sections, we summarize the properties of these
hypernovae as derived from optical light curves and spectra.  We then
show that nucleosynthesis in hypernovae is quite distinct from the
case of ordinary supernovae, thus making a unique contribution to
galactic chemical evolution.

\begin{figure}
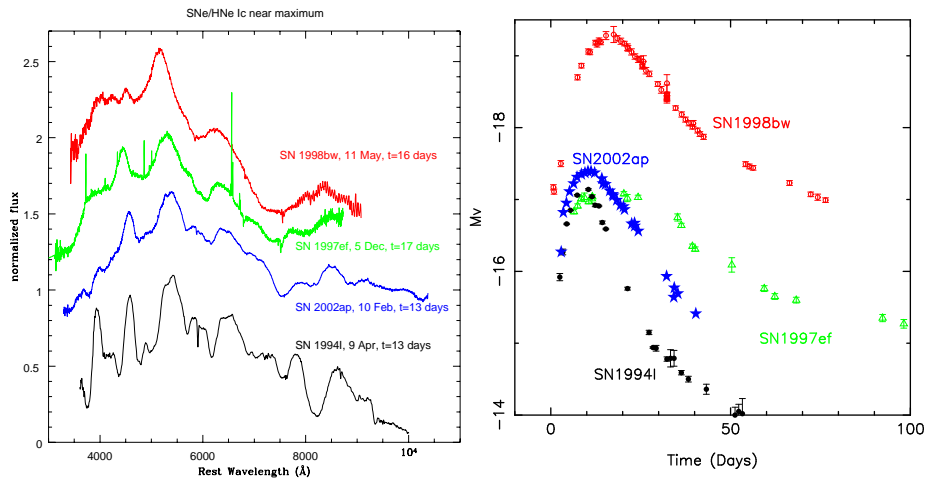

 \begin{center}
\begin{minipage}[t]{0.45\textwidth}
	\plotone{f1s02ap.epsi}
\end{minipage}
\begin{minipage}[t]{0.45\textwidth}
	\plotone{f2s02ap.epsi}
\end{minipage}
 \end{center}
 \caption{Left: The near-maximum spectra of Type Ic SNe and
 hypernovae: SNe 1998bw, 1997ef, 2002ap, and 1994I. Right: 
The observed $V$-band light curves of SNe 1998bw ({\em open
circles}), 1997ef ({\em open triangles}), 2002ap ({\em stars}), and
1994I ({\em filled circles}) (Mazzali et al. 2002). }
\label{fig:02ap}
\end{figure}

\section{Hypernova Branch and Faint Supernova Branch}

Figure 2 shows $E$ and the mass of $^{56}$Ni ejected $M(^{56}$Ni) as a
function of the main-sequence mass $M_{\rm ms}$ of the progenitor star
obtained from fitting the optical light curves and spectra.  The
estimated masses are $M_{\rm ms}$ $\gsim 60$ \Msun\ for SN1999as,
$\sim 40$ \Msun\ for SN1998bw, $\sim 35$ \Msun\ for SN1997ef, and
$\sim 20-25$ \Msun\ for SN2002ap.  These mass estimates place
hypernovae at the high-mass end of SN progenitors, as they are
consistently larger than the mass of the progenitors of normal
core-collapse SNe ($\sim 10-20$ \Msun).  Our analysis of these objects
suggests that the KE may be related to $M_{\rm ms}$.  $M(^{56}$Ni)
also appears to increase with increasing $M_{\rm ms}$, which is
important to know for the study of the chemical evolution of galaxies.

In contrast, SNe II 1997D and 1999br were very faint SNe with very low
$KE$ (Turatto et al. 1998; Hamuy 2002; Zampieri et al. 2002).  In
Figure 2, therefore, we propose that SNe from stars with $M_{\rm ms}
\gsim$ 20-25 \Msun\ have different $E$ and $M(^{56}$Ni,) with a
bright, energetic ``hypernova branch'' at one extreme and a faint,
low-energy SN branch at the other.  For the faint SNe, the explosion
energy was so small that most \Nifs\ fell back onto the compact
remnant.  Thus the faint SN branch may become a ``failed'' SN branch
at larger $M_{\rm ms}$.  Between the two branches, there may be a
variety of SNe (Hamuy 2002).

This trend might be interpreted as follows.  Stars with $M_{\rm ms}
\lsim$ 20-25 \ms\ form a neutron star, producing $\sim$ 0.08 $\pm$
0.03 \ms\ \Nifs\ as in SNe 1993J, 1994I, and 1987A
(SN 1987A may be a borderline case between the neutron star
and black hole formation).  
Stars with $M_{\rm ms} \gsim$ 20-25 \ms\
form a black hole; whether they become hypernovae or faint SNe may
depend on the angular momentum in the collapsing core, which in turn
depends on the stellar winds, metallicity, magnetic fields, and
binarity.  Hypernovae might have rapidly rotating cores owing possibly
to the spiraling-in of a companion star in a binary system.

\begin{figure}
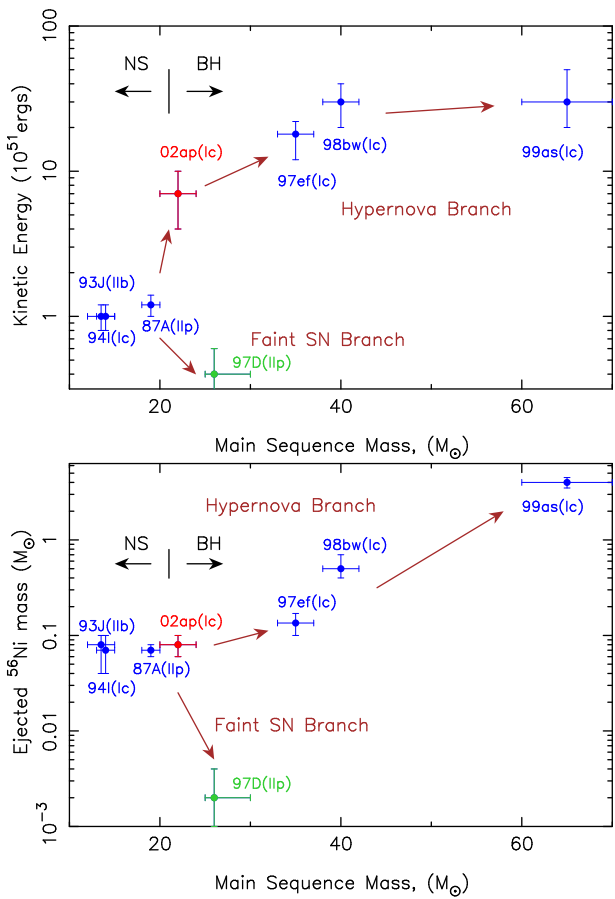

\begin{center}
  \begin{minipage}[t]{0.55\textwidth}
     \plotone{e-m.epsi}
  \end{minipage}\\
  \begin{minipage}[t]{0.55\textwidth}
     \plotone{ni-m.epsi}
  \end{minipage}
\end{center}
\caption[]{
The explosion energy and the ejected $^{56}$Ni mass
as a function of the main
sequence mass of the progenitors for several supernovae/hypernovae.
\label{fig:e-m}}
\end{figure}

\section{Aspherical Hypernova Models}

Although all modeling presented above was performed assuming spherical
symmetry, the data show evidence for significant deviations from
sphericity. In particular, some polarization was detected in both SNe
1998bw (Iwamoto et al. 1998; Patat et al. 2001) and 2002ap (Kawabata
et al. 2002; Leonard et al. 2002; Wang et al. 2002).  Furthermore, in
the case of SN~1998bw, the late time spectra showed peculiar nebular
line profiles, where the OI] 6300\AA\ line is significantly narrower
than the FeII] blend near 5200\AA\ (Patat et al. 2001).

In spherically symmetric models this is not expected, as O should
always be located above Fe in velocity space. Moreover, the OI] line
declines more slowly than the FeII] ones, possibly signalling
deposition of $\gamma$-rays in a slowly-moving, O-dominated region
(Mazzali et al 2001).  Another peculiarity is observed in SN~1997ef,
where the photosphere persists to advanced epochs, showing line
absorption at velocities of $\sim$ 2000 km s$^{-1}$, which is well
below the expected position of the mass-cut in spherically symmetric
models (Mazzali et al. 2000).  Finally, all three hypernovae show a
late decline of the light curve at epochs of a few months.

Maeda et al. (2002) calculated the nucleosynthesis in aspherical
explosions. In such a model, $^{56}$Ni is synthesized preferentially
along the polar axis, where the KE is larger, while a lot of unburned
material, dominated by O, is left at low velocity in the equatorial
region, where burning is much less efficient. A model where the ratio
of the polar-to-equatorial kinetic energy is about 8 yields an
asymmetric explosion whose properties at late times are consistent
with the observed lines of SN~1998bw if it is viewed at an angle of
about 15 degrees from the polar axis. At such an angle one might
expect that the GRB is weaker than it would be if observed along the
jet axis.  The actual aspect ratio of the ejecta is much smaller than
8:1, however, as the jet expands laterally, and this may be consistent
with the observed polarization.

The asymmetric model has a smaller total kinetic energy than the
corresponding symmetric model, as the KE away from the line of sight
is significantly reduced.  The estimate, however, is still large,
$E_{51} \sim$ 10.  The estimate of $M(^{56}$Ni) $\sim 0.6 M_\odot$
from the nebula spectra does not much depend on the asphericity
either.

\begin{figure}[t]
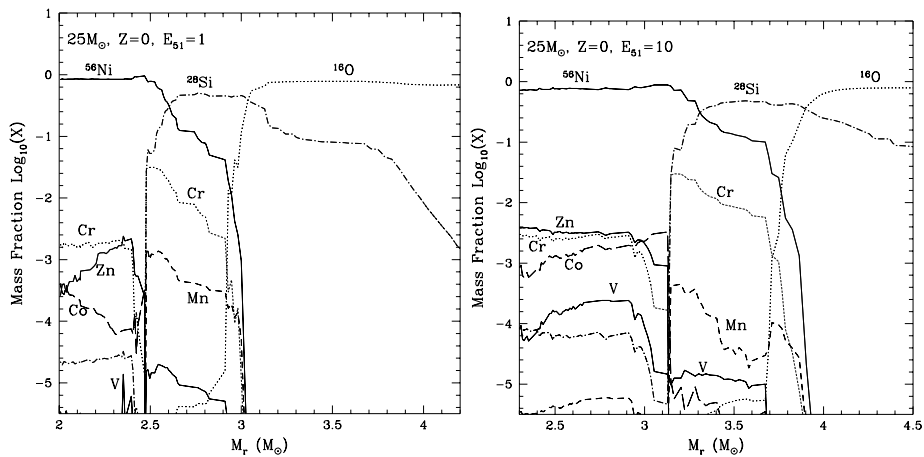

 \begin{center}
\begin{minipage}[t]{0.45\textwidth}
	\plotone{fig3a.epsi}
\end{minipage}
\begin{minipage}[t]{0.45\textwidth}
	\plotone{fig3b.epsi}
\end{minipage}
 \end{center}
 \caption[]{Abundance distribution plotted against the enclosed mass
$M_r$ after the explosion of Pop III 25 \ms\ stars with $E_{51} = 1$
(left) and $E_{51} = 10$ (right) (Umeda \& Nomoto 2002a). }
\label{fig:hnus}
\end{figure}

\section{Nucleosynthesis in Hypernova Explosions}

In core-collapse supernovae/hypernovae, stellar material undergoes
shock heating and subsequent explosive nucleosynthesis. Iron-peak
elements are produced in two distinct regions, which are characterized
by the peak temperature, $T_{\rm peak}$, of the shocked material.  For
$T_{\rm peak} > 5\times 10^9$K, material undergoes complete Si burning
whose products include Co, Zn, V, and some Cr after radioactive
decays.  For $4\times 10^9$K $<T_{\rm peak} < 5\times 10^9$K,
incomplete Si burning takes place and its after decay products include
Cr and Mn (e.g., Hashimoto, Nomoto, \& Shigeyama 1989; Thielemann,
Nomoto, \& Hashimoto 1996).

The right panel of Figure 3 shows the composition in the ejecta of a
25 \ms\ HN model ($E_{51} = 10$).  The nucleosynthesis in a normal 25
\ms\ SN model ($E_{51} = 1$) is also shown for comparison in the left
panel of Figure 3.

We note the following characteristics of nucleosynthesis with very
large explosion energies (Nomoto et al. 2001a,b):

(1) Both complete and incomplete Si-burning regions shift outward in
mass compared with normal supernovae, so that the mass ratio between
the complete and incomplete Si-burning regions becomes larger.  As a
result, higher energy explosions tend to produce larger [(Zn, Co,
V)/Fe] and smaller [(Mn, Cr)/Fe].  The elements synthesized in this
region such as $^{56}$Ni, $^{59}$Cu, $^{63}$Zn, and $^{64}$Ge (which
decay into $^{56}$Fe, $^{59}$Co, $^{63}$Cu, and $^{64}$Zn,
respectively) are ejected more abundantly than in normal supernovae.

(2) In the complete Si-burning region of hypernovae, elements produced
by $\alpha$-rich freezeout are enhanced because nucleosynthesis
proceeds at lower densities (i.e., higher entropy) and thus a larger
amount of $^{4}$He is left.  Hence, elements synthesized through
capturing of $\alpha$-particles, such as $^{44}$Ti, $^{48}$Cr, and
$^{64}$Ge (decaying into $^{44}$Ca, $^{48}$Ti, and $^{64}$Zn,
respectively) are more abundant.

(3) Oxygen burning takes place in more extended regions for the larger
KE.  Then more O, C, Al are burned to produce a larger amount of
burning products such as Si, S, and Ar.  Therefore, hypernova
nucleosynthesis is characterized by large abundance ratios of [Si/O],
[S/O], [Ti/O], and [Ca/O].

\section{Hypernovae and Galactic Chemical Evolution}

Hypernova nucleosynthesis may have made an important contribution to
Galactic chemical evolution.  In the early galactic epoch when the
galaxy was not yet chemically well-mixed, [Fe/H] may well be
determined by mostly a single SN event (Audouze \& Silk 1995). The
formation of metal-poor stars is supposed to be driven by a supernova
shock, so that [Fe/H] is determined by the ejected Fe mass and the
amount of circumstellar hydrogen swept-up by the shock wave (Ryan et
al.  1996).  Then, hypernovae with larger $E$ are likely to induce the
formation of stars with smaller [Fe/H], because the mass of
interstellar hydrogen swept up by a hypernova is roughly proportional
to $E$ (Ryan et al. 1996; Shigeyama \& Tsujimoto 1998) and the ratio
of the ejected iron mass to $E$ is smaller for hypernovae than for
normal supernovae.

\subsection {Zn, Co, Mn, Cr}

The observed abundances of metal-poor halo stars show quite interesting
pattern.  There are significant differences between the abundance
patterns in the iron-peak elements below and above [Fe/H]$ \sim -2.5$ -
$-3$.

(1) For [Fe/H]$\lsim -2.5$, the mean values of [Cr/Fe] and [Mn/Fe]
decrease toward smaller metallicity, while [Co/Fe] increases
(Fig. 4; McWilliam et al. 1995; Ryan et al. 1996).

(2) [Zn/Fe]$ \sim 0$ for [Fe/H] $\simeq -3$ to $0$ (Sneden et
al. 1991), while at [Fe/H] $< -3.3$, [Zn/Fe] increases toward smaller
metallicity (Fig. 4; Primas et al. 2000; Blake et al. 2001).

These trends cannot be explained with the conventional chemical
evolution model that uses previous nucleosynthesis yields.

The larger [(Zn, Co)/Fe] and smaller [(Mn, Cr)/Fe] in the supernova
ejecta can be realized if the mass ratio between the complete Si burning
region and the incomplete Si burning region is larger, or equivalently
if deep material from the complete Si-burning region is ejected by mixing or
aspherical effects.  This can be realized if (1) the mass cut between
the ejecta and the compact remnant is located at smaller $M_r$
(Nakamura et al. 1999), (2) $E$ is larger to move the outer edge of the
complete Si burning region to larger $M_r$ (Nakamura et al. 2001), or
(3) asphericity in the explosion is larger.

Among these possibilities, a large explosion energy $E$ enhances
$\alpha$-rich freezeout, which results in an increase of the local
mass fractions of Zn and Co, while Cr and Mn are not enhanced (Umeda
\& Nomoto 2002a,b).  Models with $E_{51} = 1 $ do not produce
sufficiently large [Zn/Fe].  To be compatible with the observations of
[Zn/Fe] $\sim 0.5$, the explosion energy must be much larger, i.e.,
$E_{51} \gsim 20$ for $M \gsim 20 M_\odot$, i.e., hypernova-like
explosions of massive stars ($M \gsim 25 M_\odot$) with $E_{51} > 10$
are responsible for the production of Zn.

In the hypernova models, the overproduction of Ni, as found in the
simple ``deep'' mass-cut model, can be avoided.  Therefore, if
hypernovae made significant contributions to the early Galactic
chemical evolution, it could explain the large Zn and Co abundances
and the small Mn and Cr abundances observed in very metal-poor stars
as seen in Figure 4.

\begin{figure}[t]
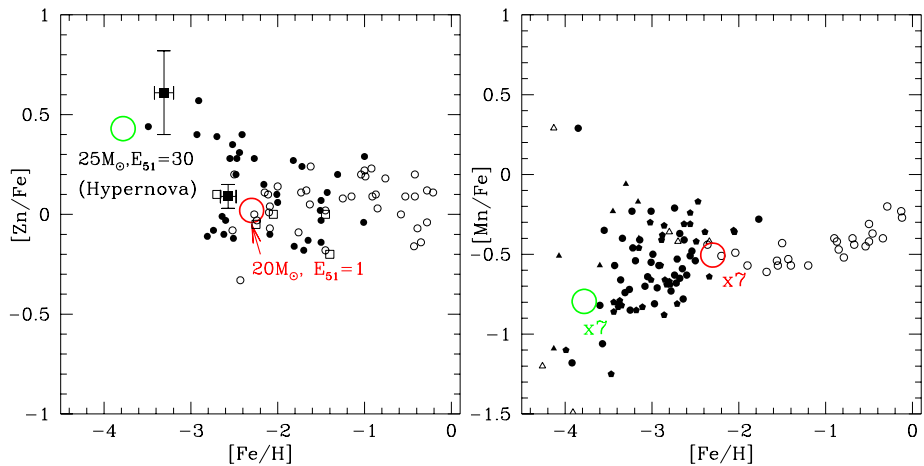

  \begin{center}
  \begin{minipage}[t]{0.45\textwidth}
	\plotone{fig4a.epsi}
  \end{minipage}
  \begin{minipage}[t]{0.45\textwidth}
	\plotone{fig4b.epsi}
  \end{minipage}  \end{center}
\caption{
Observed abundance ratios of [Zn/Fe] and [Mn/Fe] and the theoretical
abundance patterns for a normal SN II (20$M_\odot$, $E_{51}=1$) and a
hypernova (25$M_\odot$, $E_{51}=30$) models (Umeda \& Nomoto 2002b).
\label{fig:znfe}}
\end{figure}

\subsection{Fe, Ti}

The Fe mass observed in hypernovae show a trend with the progenitor
mass, ranging from $\sim$ 5 \ms\ in SN 1999as to 0.07 \ms\ in SN
2002ap.  Thus [O/Fe] in the ejecta of most hypernovae may be larger
than the solar ratio (see Umeda \& Nomoto 2002a).  The small [O/Fe]
observed in some metal-poor stars and galaxies might be the results of
SNe from 13 - 15 \ms\ stars (Nomoto et al. 1997) or possibly very
massive hypernovae rather than Type Ia supernovae.  In contrast,
[O/Fe] must be very large in the faint SN branch.  Therefore, the
scatter of [O/Fe] in metal-poor stars might provide constraints on the
fraction of these branches (e.g., Argast et al. 2002).

It has been pointed out that Ti is deficient in Galactic chemical
evolution models using supernova yields currently available (e.g.,
Timmes et al. 1996; Thielemann et al. 1996), especially at [Fe/H]
$\lsim -1$ when SNe Ia have not contributed to the chemical evolution.
However, if the contribution from hypernovae to Galactic chemical
evolution is relatively large (or supernovae are more energetic than
the typical value of $E_{51}=1$), this problem could be relaxed.  The
$\alpha$-rich freezeout is enhanced in hypernovae, so that $^{48}$Ti
could be ejected more abundantly.

\section {Starburst Galaxy M82 and Hypernovae}

X-ray emissions from the starburst galaxy M82 were observed with ASCA
and the abundances of several heavy elements were obtained (Tsuru et
al. 1997).  Tsuru et al. (1997) found that the overall metallicity of
M82 is quite low, i.e., O/H and Fe/H are only 0.06 - 0.05 times solar,
while Si/H and S/H are $\sim$ 0.40 - 0.47 times solar.  This implies
that the abundance ratios are peculiar, i.e., the ratio O/Fe is about
solar, while the ratios of Si and S relative to O and Fe are as high
as $\sim$ 6 - 8.  These ratios are very different from those ratios in
SNe II.  Compared with normal SNe II, the important characteristic of
hypernova nucleosynthesis is the large Si/O, S/O, and Fe/O ratios.
The good agreement between the hypernova model ($E_{51}=$ 30) and the
observed abundances in M82 is seen in Umeda et al. (2002).

Hypernovae could also produce larger $E$ per oxygen mass than normal
SNe II, as required for M82.  We therefore suggest that hypernova
explosions may make important contributions to the metal enrichment
and energy input to the interstellar matter in M82.  The age of
starburst activity is estimated to be $lsim 10^7$ years (Strickland
2002), which is so young that only massive stars ($M >$ 25 $M_\odot$)
contributed to nucleosynthesis in M82.

\section{Concluding Remarks}

We have shown that signatures of hypernova nucleosynthesis are seen in
the abundance patterns in very metal poor stars and the starburst
galaxy M82.  (See also the abundance pattern in X-ray Nova Sco;
Israelian et al. 1999; Podsiadlowski et al. 2002).  We suggest that
hypernovae of massive stars may make important contributions to the
Galactic (and cosmic) chemical evolution, especially in the early low
metallicity phase.  The IMF of Pop III stars might be different from
that of Pop I and II stars, and that more massive stars are abundant
for Pop III.

%


\begin{references}

\reference Argast, D., Samland, M., Thielemann, F.-K., \& Gerhard,
O.E. 2002, A\&A, 388, 842

\reference Audouze, J., \& Silk, J. 1995, ApJ, 451, L49

\reference Blake, L.A.J., Ryan, S.G., Norris, J.E., \& Beers, T.C. 2001, 
Nucl. Phys. A., 688, 502

\reference Galama, T., et al. 1998, Nature, 395, 670

\reference Germany L.M., et al. 2000, ApJ, 533, 320

\reference Hamuy, M. 2002, ApJ, submitted

\reference Hashimoto, M., Nomoto, K., \& Shigeyama, T. 1989, A\&A, 210, L5

\reference Hatano, K., Branch, D., Nomoto, K., et al. 2001, BAAS, 198, 3902

\reference Israelian, G., Rebolo, R., Basri, G., et al. 1999, Nature, 401, 142

\reference Iwamoto, K., Mazzali, P.A., Nomoto, K., et al. 1998, 
Nature, 395, 672

\reference Iwamoto, K., Nakamura, T., Nomoto, K., et al. 2000, ApJ, 534, 660

\reference Kawabata, K., et al. 2002, ApJ, submitted (astro-ph/0205014)

\reference Knop, R., Aldering, G., Deustua, S., et al., 1999, IAU
Circ., 7128

\reference Leonard, D.C., et al. 2002, PASP, submitted (astro-ph/0206368)

\reference Maeda, K., Nakamura, T., Nomoto, K., et al. 2002, ApJ, 565, 405

\reference Matheson, T., Filippenko, A.V., Li, W., et al.  2001, AJ,
121, 1648

\reference Mazzali, P.A., Iwamoto, K., \& Nomoto, K. 2000, ApJ, 545, 407

\reference Mazzali, P.A., Nomoto, K., Patat, F., \& Maeda, K. 2001, 
ApJ, 559, 1047

\reference Mazzali, P.A., Deng, J., Maeda, K., Nomoto, K., et al.,
2002, ApJ, 572, L61

\reference McWilliam, A., Preston, G.W., Sneden, C., \& Searle,
L. 1995, AJ, 109, 2757

\reference Nakamura, T., Umeda, H., Iwamoto, K., Nomoto, K., Hashimoto, 
M., Hix, R.W., Thielemann, F.-K. 2001, ApJ, 555, 880

\reference Nakamura, T., Umeda, H., Nomoto, K., Thielemann, F.-K., \&
Burrows, A. 1999, ApJ, 517, 193

\reference Nomoto, K., Hashimoto, M, Tusjimoto, T., et al. 1997,
Nucl. Phys. A616, 79c

\reference Nomoto, K., Mazzali, P., Nakamura, T., et al. 2001a, in
{\em Supernovae and Gamma Ray Bursts}, eds. M. Livio, et
al. (Cambridge Univ. Press) 144 (astro-ph/0003077)

\reference Nomoto, K., Maeda, K., Umeda, H., \& Nakamura, T. 2001b, 
in {\em The Influence of Binaries on Stellar Populations Studies},
ed. D. Vanbeveren (Kluwer) 507 (astro-ph/0105127)

\reference Patat, F., et al. 2001, ApJ, 555, 900

\reference Podsiadlowski, Ph., Nomoto, K., Maeda, K., Nakamura, T., Mazzali,
P.A., \& Schmidt, B. 2002, ApJ, 567, 491

\reference Primas, F., Brugamyer, E., Sneden, C., et al. 2000, in {\em
The First Stars}, ed. A. Weiss, et al. (Springer), 51

\reference Rigon, L., Turatto, M., Benetti, S., et al. 2002, MNRAS,
submitted

\reference Ryan, S.G., Norris, J.E. \& Beers, T.C. 1996, ApJ, 471, 254

\reference Shigeyama. T., \& Tsujimoto, T. 1998, ApJ, 507, L135

\reference Sneden, C., Gratton, R.G., \& Crocker, D.A. 1991, A\&A, 246, 354

\reference Strickland, D. 2002, in {\em Chemical Enrichment of
Intracluster and Intergalactic Medium}, ASP Conference Series 253, 
ed. R. Fusco-Feminao, \& Matteucci (ASP), 387

\reference Thielemann, F.-K., Nomoto, K., \& Hashimoto, M. 1996, ApJ, 460, 408

\reference Timmes, F.X., Woosley, S.E., Hartmann, D.H., Hoffman, R.D. 1996, 
ApJ, 464, 332

\reference Tsuru, T. G., Awaki, H., Koyama K., Ptak, A. 1997, PASJ, 49, 619

\reference Turatto, M., Mazzali, P.A., Young, T., Nomoto, K., et al.,
1998, ApJ, 498, L129

\reference Turatto, M., Suzuki, T., Mazzali, P.A., et al., 2000, ApJ, 534, L57

\reference Umeda, H., \& Nomoto, K. 2002a, ApJ, 565, 385

\reference Umeda, H., \& Nomoto, K. 2002b, in Nuclear Astrophysics, ed.
W. Hillebrandt \& E. M\"ller (Garching: MPA), 164 (astro-ph/0205365)

\reference Umeda, H., Nomoto, K., Tsuru, T., \& Matsumoto, H, 2002, ApJ, 
578, in press (astro-ph/0207067) %

\reference Wang, L. et al. 2002, ApJ, submitted (astro-ph/0206386)

\reference Woosley, S.E., Eastman, R.G., \& Schmidt, B.P. 1999, ApJ, 516, 788

\reference Zampieri, L., Pastorello, A, Turatto, M., et al. 2002,
MNRAS, submitted

\end{references}
\end{document}